\documentclass[
reprint,
longbibliography,
amsmath, 
amssymb, 
aps, 
superscriptaddress,
pra,
]
{revtex4-2}

\usepackage{graphicx}
\usepackage{dcolumn}
\usepackage{bm}
\usepackage{braket} 
\usepackage{hyperref}
\usepackage{subfigure} 
\usepackage{nccmath}
\usepackage{amsthm} 
\usepackage{mathrsfs} 
\usepackage{xfrac} 
\usepackage{pifont}
\usepackage[export]{adjustbox}
\usepackage{bbm} 
\usepackage[utf8]{inputenc} 
\usepackage[normalem]{ulem}
\usepackage[usenames,dvipsnames,svgnames,table]{xcolor}
\usepackage[sort&compress]{natbib}
\usepackage{color}
\usepackage{longtable}
 
\usepackage{commath}
\usepackage{multirow}
\usepackage{indentfirst}

\begin{document}
\title{Transfer learning of optimal QAOA parameters in combinatorial optimization}

\author{J. A. Monta\~nez-Barrera}
 	\altaffiliation{Corresponding author: J. A. Monta\~nez-Barrera; j.montanez-barrera@fz-juelich.de}
	\affiliation{Jülich Supercomputing Centre, Institute for Advanced Simulation, Forschungszentrum Jülich, 52425 Jülich, Germany}
\author{Dennis Willsch}
	\affiliation{Jülich Supercomputing Centre, Institute for Advanced Simulation, Forschungszentrum Jülich, 52425 Jülich, Germany}

     \affiliation{Faculty of Medical Engineering and Technomathematics, University of
    Applied Sciences Aachen, 52428 Jülich, Germany}
\author{Kristel Michielsen}
	\affiliation{Jülich Supercomputing Centre, Institute for Advanced Simulation, Forschungszentrum Jülich, 52425 Jülich, Germany}
	\affiliation{AIDAS, 52425 Jülich, Germany}
	\affiliation{RWTH Aachen University, 52056 Aachen, Germany}
\begin{abstract}

Solving combinatorial optimization problems (COPs) is a promising application of quantum computation, with the Quantum Approximate Optimization Algorithm (QAOA) being one of the most studied quantum algorithms for solving them. However, multiple factors make the parameter search of the QAOA a hard optimization problem. In this work, we study transfer learning (TL), a methodology to reuse pre-trained QAOA parameters of one problem instance into different COP instances. This methodology can be used to alleviate the necessity of classical optimization to find good parameters for individual problems. To this end, we select small cases of the traveling salesman problem (TSP), the bin packing problem (BPP), the knapsack problem (KP), the weighted maximum cut (MaxCut) problem, the maximal  independent set (MIS) problem, and portfolio optimization (PO), and find optimal $\beta$ and $\gamma$ parameters for $p$ layers. We compare how well the parameters found for one problem adapt to the others. Among the different problems, BPP is the one that produces the best transferable parameters, maintaining the probability of finding the optimal solution above a quadratic speedup over random guessing for problem sizes up to 42 qubits and $p = 10$ layers. 
Using the BPP parameters, we perform experiments on IonQ Harmony and Aria, Rigetti Aspen-M-3, and IBM Brisbane of MIS instances for up to 18 qubits. The results indicate that IonQ Aria yields the best overlap with the ideal probability distribution. Additionally, we show that cross-platform TL is possible using the D-Wave Advantage quantum annealer with the parameters found for BPP. We show an improvement in performance compared to the default protocols for MIS with up to 170 qubits. Our results suggest that there are QAOA parameters that generalize well for different COPs and annealing protocols.

\begin{description}
	\vspace{0.2cm}
	\item[Keywords] QUBO; transfer learning; knapsack; bin packing; portfolio optimization; TSP;
    maxcut;
    MIS; quantum optimization; QAOA; combinatorial optimization.
\end{description}

\end{abstract}

\maketitle

\section{\label{int}Introduction}

Solving COPs is perceived as one of the major applications for the near future of quantum computation. There are three main reasons for this. First, COPs can be effectively encoded in Hamiltonians, where the ground state corresponds to the optimal solution of the problem \cite{Lucas2014, Kochenberger2014}. Second, COPs have practical applications and are hard to solve \cite{Ohzeki2020}. Third, quantum algorithms to solve these problems need few resources and can be tested on current state-of-the-art quantum hardware \cite{Harrigan2021, Niroula2022, Shaydulin2023}.

One of the most studied quantum algorithms for solving COPs is QAOA \cite{Farhi2014}. QAOA consists of $p$ layers, each of which includes the COP cost Hamiltonian encoded in a parametric unitary gate with parameters $\gamma_i$ and a ``mixer'' parametric unitary gate with parameters $\beta_i$. In this context, the parameters are adjusted to improve the probability of finding good solutions to the problem. Different techniques have been proposed to improve the performance of QAOA. For example, warm-start \cite{Egger2021, Tate2023, Tate2023b} is used
to adjust QAOA in order to start it from solutions obtained by classical algorithms or recursive QAOA \cite{FinZgar2024} which uses the QAOA to produce sequences of reduced problems. In \cite{Abbas2023, He2024} are reviews of different techniques used in QAOA. To some extent, QAOA can be seen as a trotterized quantum annealing protocol where the number of layers $p$ determines the precision of the solution \cite{Cerezo2021, Willsch2022}. 

The most used technique to find the $\gamma$ and $\beta$ parameters in QAOA consists of a subroutine of classical optimization where QAOA provides samples used to obtain the expectation value of the cost Hamiltonian. Subsequently, a classical solver produces new $\gamma$ and $\beta$ parameters to minimize the expectation value. This process is repeated until a good set of parameters is found or other stopping criteria is reached. There are many limitations to this approach, e.g., finding the optimal set ofparameters has been proved to be an NP-Hard problem \cite{Bittel2021, Akshay2020QAOAReachabilityDeficitBarrenPlateauQAOA}. Additionally, the energy landscape of the parameterized QAOA has many local minima where classical solvers can easily get trapped \cite{Bittel2021, Kremenetski2021}. Moreover, implementations on real hardware face an even greater challenge, the noise inherent in current quantum devices makes the search for the minima of the objective function unfeasible after only a few QAOA layers \cite{Zhou2020, Kossmann2022}.

An alternative strategy that partially overcomes these problems and has gained attention recently is the use of transferred parameters \cite{Shaydulin2023b, Basso2022, Boulebnane2022}. The idea is to find a good set of parameters using the classical optimization loop for small instances of a problem, where it is easier to find good parameters, and then, reuse them on larger instances. For instance, in  \cite{Brandao2018}, TL of 3-regular graphs in the MaxCut problem was initially introduced as an alternative to do extensive classical optimization for the QAOA parameters. Subsequently, \cite{Lotshaw2021} extended this concept to non-isomorphic unweighted graphs in the MaxCut problem and showed optimal QAOA parameter concentration for large graphs. In \cite{Galda2021}, different properties of random graphs in the MaxCut problem were identified as indicative characteristics of parameter transferability. In \cite{Wurtz2021}, numerical evidence supporting an approximation ratio exceeding the worst-case scenario by Goemans-Williamson (GW) was presented for parameters transferred in 3-regular graphs of MaxCut, specifically for $p < 12$. Furthermore, \cite{Shaydulin2023} provided indications of TL capabilities in the context of the weighted MaxCut problem. However, the prospects of TL between different COPs have not been systematically investigated to this point.

In this paper, we extend the study of TL capabilities of optimal QAOA parameters across different COPs. This is motivated by the fact that finding a good set of parameters that works for different COPs, independent of their construction or problem size, can be used in larger problems or as an initial guess for doing a subsequent classical optimization on specific problem instances. This initial guess is therefore not biased by the given problem class or specific instance. To this end, we select random instances of TSP, KP, BPP, PO, MaxCut, and MIS. We first use the Constrained Optimization BY Linear Approximation (COBYLA) \cite{Powell1994COBYLA} optimization method to find $\beta$ and $\gamma$ parameters for problems with up to 20 qubits. 

We study the transfer of those parameters to (i) the same COP with up to 42 qubits and (ii) other random instances of completely different COPs. We thereby demonstrate that parameters can not only be transferred to larger instances of the same problem but also to completely different problems. We use the probability of finding the ground state, $probability(^*x)$, as the performance metric of the TL methodology. Of all the COPs studied, BPP is the one that shows the best TL capabilities. 

Furthermore, we study the practical TL performance on various quantum technology platforms, namely Aspen-M3 from Rigetti, Harmony and Aria from IonQ, and ibm\_brisbane from IBM with problem sizes 8, 14, and 18 qubits. We present solutions for the MIS using QAOA with p=10 and TL from BPP. Our results suggest that even in the 14 qubit case for QAOA with p=10, which corresponds to 640 CNOT gates, a positive net gain of TL may still be observable using IonQ Aria.

Additionally, we explore ``cross-platform TL'', by transferring the QAOA protocol to a quantum annealing protocol. To this end,  we use a D-Wave quantum annealer \cite{johnson2011quantum, King_2023}, D-Wave Advantage, with a modified custom anneal schedule that reflects the learned TL parameters to mimic the QAOA $\beta$ and $\gamma$ parameters separately. We study MIS from 100 to 170 qubits and find that the TL protocol of the $\beta$ parameters performs consistently better than the default D-Wave Advantage annealing protocol for this particular problem.

The rest of the paper is organized as follows. Section~\ref{Sec:QUBO} provides a description of the COPs used in this work, the TL methodology, the postprocessing technique for the real hardware implementation to mitigate some of the noise, and a description of the cross-platform TL approach. In Sec.~\ref{Sec:Results} we present our results and a discussion. Finally, Sec.~\ref{Sec:Conclusion} contains our conclusions. The source code for the results shown here can be found at \url{https://jugit.fz-juelich.de/qip/transfer-learning-QAOA}.

\section{Transfer learning in QAOA} \label{Sec:QUBO}

In the context of QAOA, we refer to TL as the use of pre-optimized $\gamma = [\gamma_0, \gamma_1, ..., \gamma_{p-1}]$ and $\beta = [\beta_0, \beta_1, ..., \beta_{p-1}]$ for parameters on problems that were not used for the optimization. In this methodology, the first step is to find an optimal set of parameters that works well for a specific problem. Then, we test if the optimized parameters work well on different instances of the same and other problems. 

\begin{figure*}[!tbh]
\centering
\includegraphics[width=18cm]{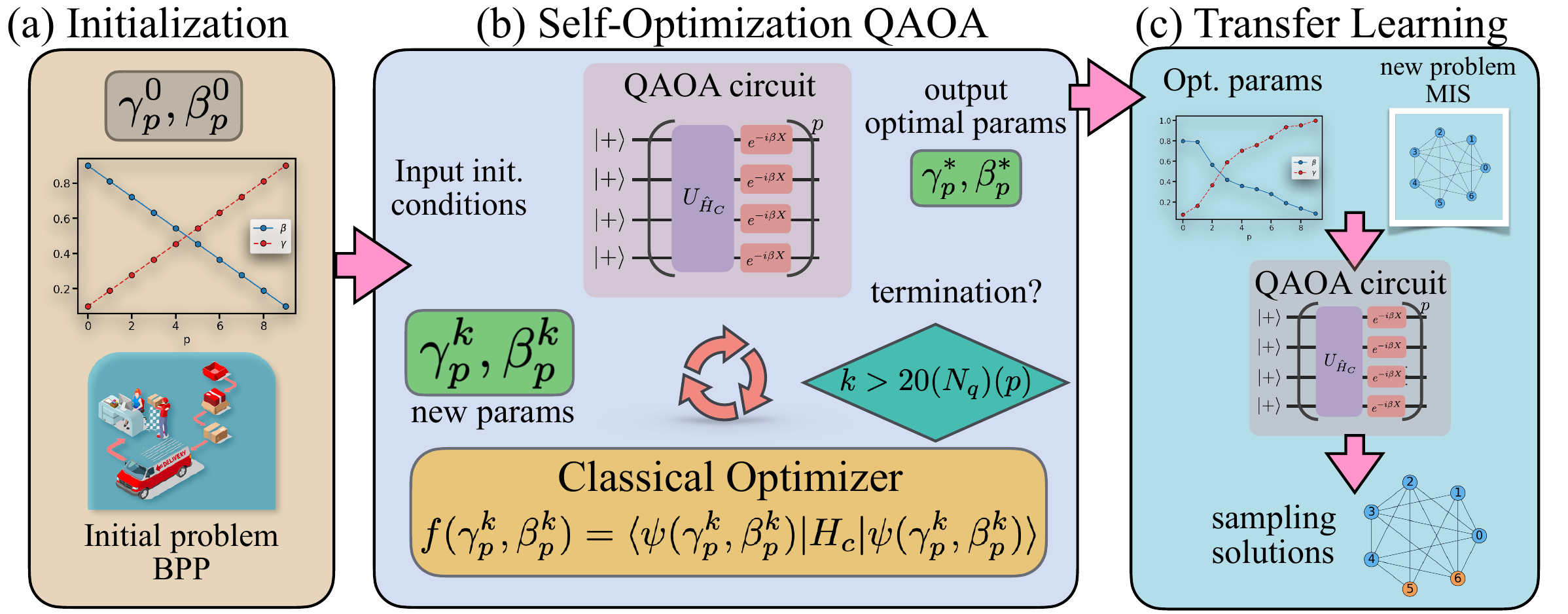}
\caption{\label{tl_method} Example of the TL methodology for transferring the parameters from BPP to MIS. (a) Quantum annealing initialization of the QAOA parameters for $p=10$ layers using the BPP, (b) self-optimization step using QAOA, (c) final $\beta$ and $\gamma$ parameters transferred to the MIS problem.}
\end{figure*}

We study random instances of TSP, BPP, MIS, KP, PO, and MaxCut using QAOA with $p=10$. We employ a quantum annealing initialization of the QAOA parameters \cite{Zhou2020, Sack2021}. To find the minimum of the cost function, COBYLA is used with a maximum number of iterations given by max\_iter = $20N_qp$, where $N_q$ is the number of qubits needed by the problem and $p$ is the number of QAOA layers. Fig.~(\ref{tl_method}) shows the methodology used for (a) the initialization of the $\gamma$ and $\beta$ parameters on the problem selected, (b) the loop of self-optimization where the parameters are updated improving the expectation value of the cost Hamiltonian of the problem, and (c) the TL of the parameters from the problem (BPP) on a new problem – in this case, the MIS. If the TL is successful, the QAOA circuit should sample good solutions for the new problem.

We pick 5 random instances for each problem size. For the TSP, we use instances with 3, 4, 5, and 6 cities (9, 16, 25, and 36 qubits), where the distances between cities are randomly chosen from a normal distribution with a mean value of 10 and a standard deviation of 0.1. This set of values is chosen because it has been shown to produce hard instances of the TSP \cite{Cheeseman1991}. In the BPP, we consider scenarios involving 3, 4, 5, and 6 items (12, 20, 30, and 42 qubits). The weight of each item is randomly chosen from integer values between 1 and 10, and 20 is the maximum weight of the bins. The MaxCut, MIS, KP, and PO problem sizes are 4, 6, 8, 10, 12, 14, 16, 18, 20, 25, 30, 35, 40, and 42. For MaxCut problems, we use randomly weighted edges with weights between 0 and 1 and a probability of having an edge between any two vertices as 70\%. In MIS problems, edges between nodes are randomly selected with a 50\% probability of having an edge. For KP problems, item values are randomly chosen from integer values between 5 to 63, weights from 1 to 20, and the maximum weight is set to half of the sum of all weights. Finally, for PO, correlation matrix values are chosen randomly from $[-0.1, 0, 0.1, 0.2]$, asset costs varying between 0.5 and 1.5, and the budget is set to half of the total assets cost. For large problems, we simulate them using JUQCS–G software \cite{Willsch2022} on JUWELS Booster, a cluster of 3744 NVIDIA A100 Tensor Core GPUs, integrated into the modular supercomputer JUWELS \cite{Krause2019, JuwelsClusterBooster}.

For the inequality constraints in the KP, PO, and BPP, we use the unbalanced penalization approach \cite{Montanez-Barrera2022, Montanez-Barrera2023}. Once the QUBO is generated with quadratic penalization terms and translated to the Ising Hamiltonian representation, the Hamiltonian is normalized in all the cases. This step is helpful to bring the problem Hamiltonian to a similar range of energy. A similar procedure is used by D-Wave annealers with the auto\_scale functionality \cite{dwave_solver_parameters}. In Appendix \ref{appendix}, we explain in detail the problems and the parameters used in this work.

To quantify the performance of TL, we use the probability of success, $probability(^*x)$. This is calculated from the final state vector $|\psi\rangle$ (after the QAOA protocol has been implemented) given by
\begin{equation}
    |\psi\rangle = \sum_{x \in \{0,1\}^n} \alpha_x |x\rangle,
\end{equation}

where $n$ is the number of qubits involved. The probability of success is calculated by

\begin{equation}
    probability(^*x) = \sum_{x_i \in ^* x} |\alpha_{x_i}|^2,
\end{equation}

where $^* x$ is the set of all the optimal solutions of the given problem. The set of $^* x$ solutions has been previously obtained using the classical solver CPLEX \cite{Bliek2014SolvingMQ}. Note that the $probability(^* x)$ is preferred here instead of the approximation ratio $r$ as it has been shown that $r$ can be a misleading metric in constrained problems \cite{Schulz2024}.

\subsection*{Mitigation: Hamming distance 1}

Our approach uses the Hamming distance 1 strategy as a post-processing technique to reduce errors in real quantum devices. This involves applying a bitflip to each position within the output bit-string, to mitigate single-qubit bitflips. The computational overhead of this post-processing method is linear,  $\mathrm{O}(N N_q)$, where N represents the number of samples and $N_q$ is the number of qubits. It is important to note that the success of this method in improving the probability of the ground state relies on the large probability of obtaining the ground state compared to the number of samples used. Also, the error must be low enough to have only a single qubit error in the samples.

\subsection*{Cross-platform TL}\label{Sec:Cross-platform}

\begin{figure}[!tbh]
\centering
\includegraphics[width=8.5cm]{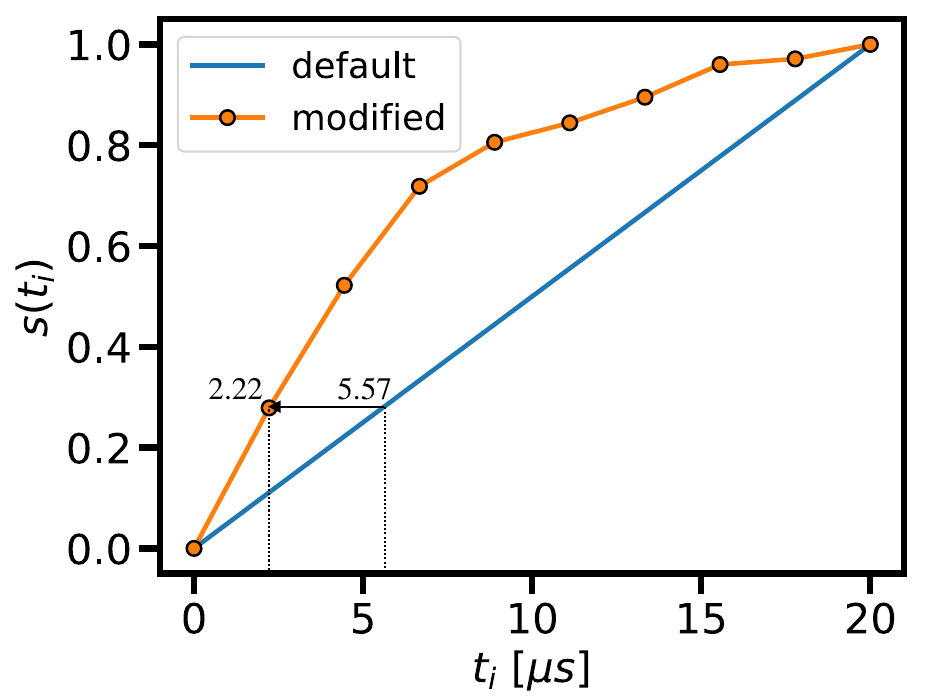}
\caption{\label{Fig:dwave} D-Wave modified schedule example. The circles represent the modified point in the realtion between $s_i$ and $t_i$.}
\end{figure}

In addition to TL between COPs, we also investigate the possibility of applying TL across platforms. In the case of quantum annealing, we want to test the capabilities to transfer the QAOA parameters for solving COPs using D-Wave Advantage. The default quantum annealing protocol on D-Wave Advantage, as proposed by Johnson et al. \cite{Johnson2010}, is represented by

\begin{equation}
H(s) = -\frac{A(s)}{2} \sum_i \sigma_x^i + \frac{B(s)}{2}\,H_c, 
\end{equation}
where $s\in[0,1]$ is a parameter that represents normalized time, $H_c$ is the problem cost Hamiltonian (see Appendix \ref{appendix} for details), $A(s)$ is the annealing protocol associated to the mixer, $B(s)$ is the annealing protocol of the cost Hamiltonian, and $\sigma_x^i$ is the Pauli-$x$ matrix for qubit $i$. The D-Wave device has hardware-dependent, fixed annealing functions $A(s)$ and $B(s)$ that cannot be changed. They are documented in spreadsheets that can be found for each device in the online D-Wave documentation. What can be changed by the user, however, is the mapping $s(t)$ between the normalized annealing fraction $s\in[0,1]$ and the real-time $t\in[500\,\mathrm{ns},2000\,\mu\mathrm{s}]$. The value of $s$ is given by $s_i = t_i/t_f$, where $t_i$ is the instantaneous real time and $t_f$ is the total annealing time. D-Wave Advantage allows modification of the annealing schedule by specifying a maximum of 12 points for the relation $s_i = f(t_i)$ between a normalized time and instantaneous real-time. This flexibility enables us to implement a custom schedule for $A(s)$ or $B(s)$. For example, Fig.~\ref{Fig:dwave} shows a new relationship between the $s(t_i)$ and the time $t_i$. Hence, what originally happened at $t=5.57 \mu s$ in the default schedule now happens at $t=2.2\mu s$ in the modified schedule.` We utilize parameters from the TL parameters of QAOA to  modify the quantum annealing schedule on a D-Wave quantum annealer. 
    
\section{Results}\label{Sec:Results}

In this section, we first study, by numerical simulation, TL from one COP to another COP. Using the best parameters, we then perform experiments on various quantum hardware devices. Finally, we investigate cross-platform TL, i.e., we learn parameters using a gate-based quantum computing model and then transfer them to a physical device implementing the quantum annealing model.

\begin{figure}[!tbh]
\centering
\includegraphics[width=8.5cm]{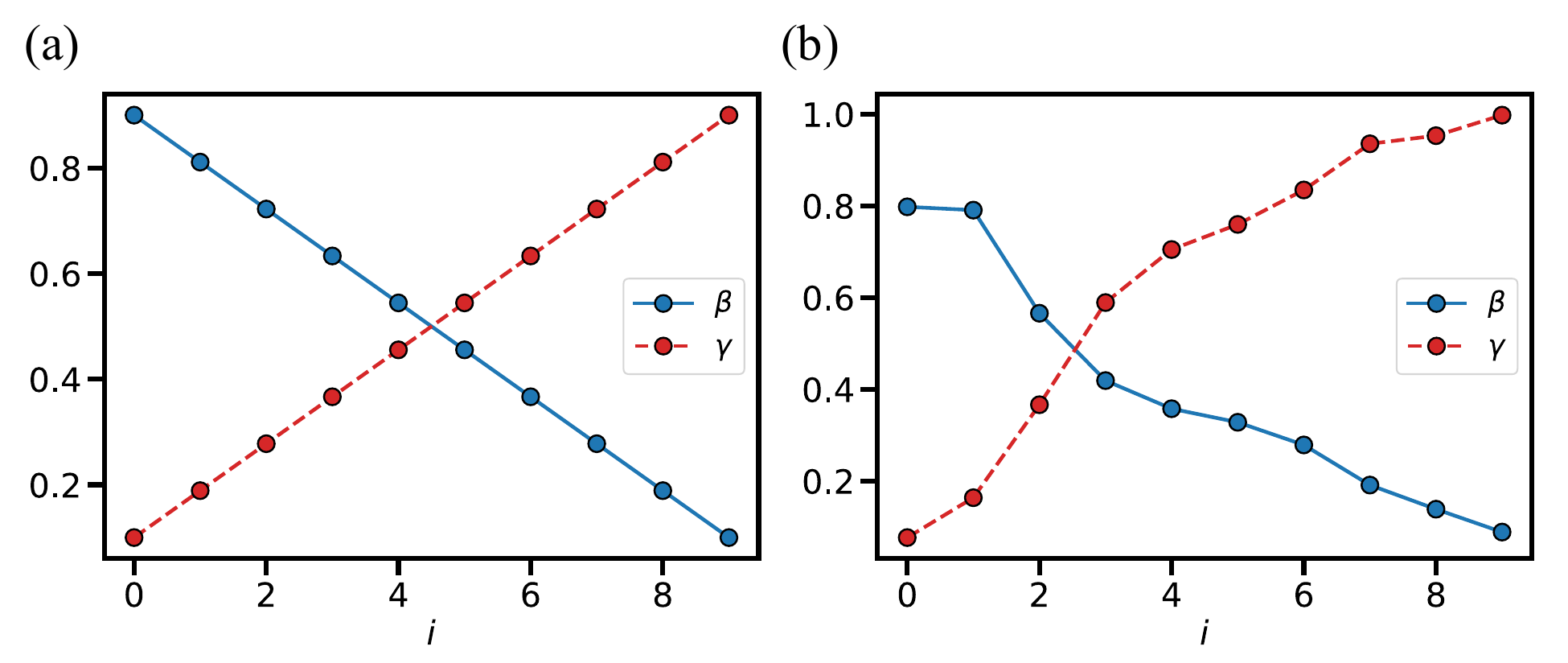}
\caption{\label{tl_qaoa} Example for the QAOA parameter optimization of the BPP. (a) Quantum annealing initialization of the QAOA parameters for $p=10$ layers, and (b) final $\beta_i$ and $\gamma_i$ for $i=0, ..., p-1$ angles for the BPP with 3 items (12 qubits).}
\end{figure}

The initialization used for all problems was a linear ramp quantum annealing scheme (see Fig.~\ref{tl_qaoa} (a)). Fig.~\ref{tl_qaoa} (b) shows the final $\gamma$ and $\beta$ parameters of a random instance of the BPP. These parameters are used in subsequent plots to show the capabilities of TL from the BPP.

\begin{figure*}[!tbh]
\centering
\includegraphics[width=18cm]{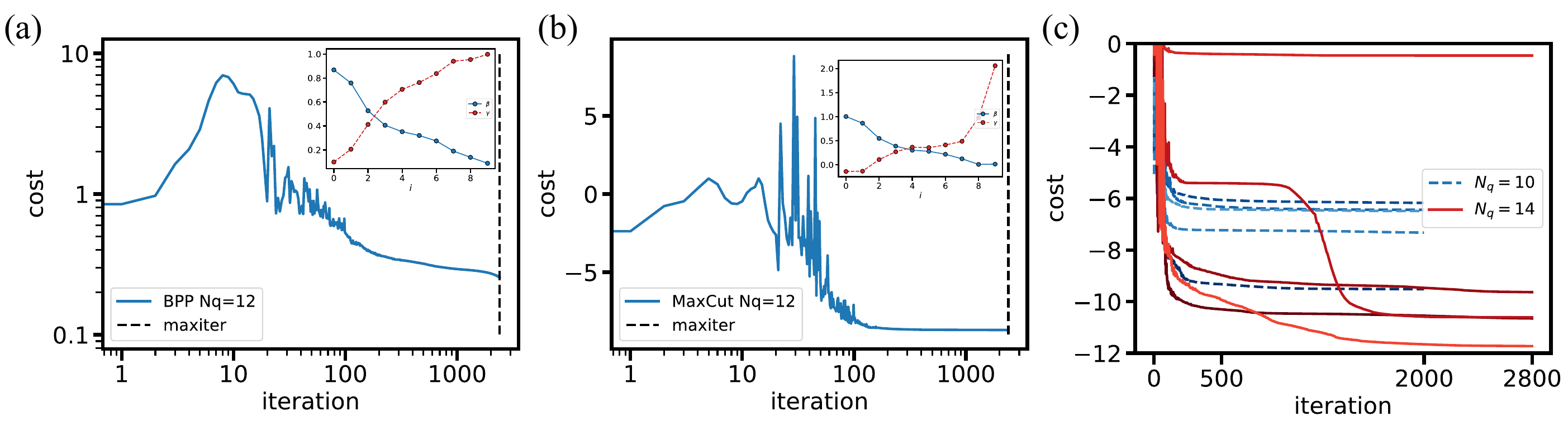}
\caption{\label{cost_func} (a)-(b)Classical optimization of the cost function value at each iteration for a random 12 qubit (3 items) BPP problem and 12 nodes MaxCut problem. (a) BPP (b) MaxCut. The inset plot shows the final parameters found for each problem. The maximum number of iterations is 2400. (c) MaxCut cost function versus the number of iterations for the classical optimization for problem sizes 10 and 14 qubits. Different lines represent the different seeds.}
\end{figure*}

Figure \ref{cost_func} shows the classical optimization evolution for (a) BPP and (b)-(c) MaxCut problem using QAOA with $p=10$. The cost represents the Hamiltonian energy at the given iteration. The optimization process effectively helps to improve the $\gamma$ and $\beta$ parameters reducing the average cost for both problems. The final schedules in the inset of Fig. \ref{cost_func}(a)-(b) seem to differ in magnitude but some characteristics from the initial annealing schedule are conserved, i.e., $\beta$ goes down to 0 and $\gamma$ goes up to some value. Fig. \ref{cost_func}-(c) shows the cost for the different seeds of problems sizes 10 and 14 qubits. For problem size 10 qubits, the optimization process seems not to have an effect after 500 iterations while for 3 out of 5 cases of the 14-qubit problems the optimization is still effective after 1000 iterations. This suggests that the number of iterations to find QAOA parameters is not only affected by $p$ but also by the number of qubits.

\begin{figure}[!tbh]
\centering
\includegraphics[width=8.5cm]{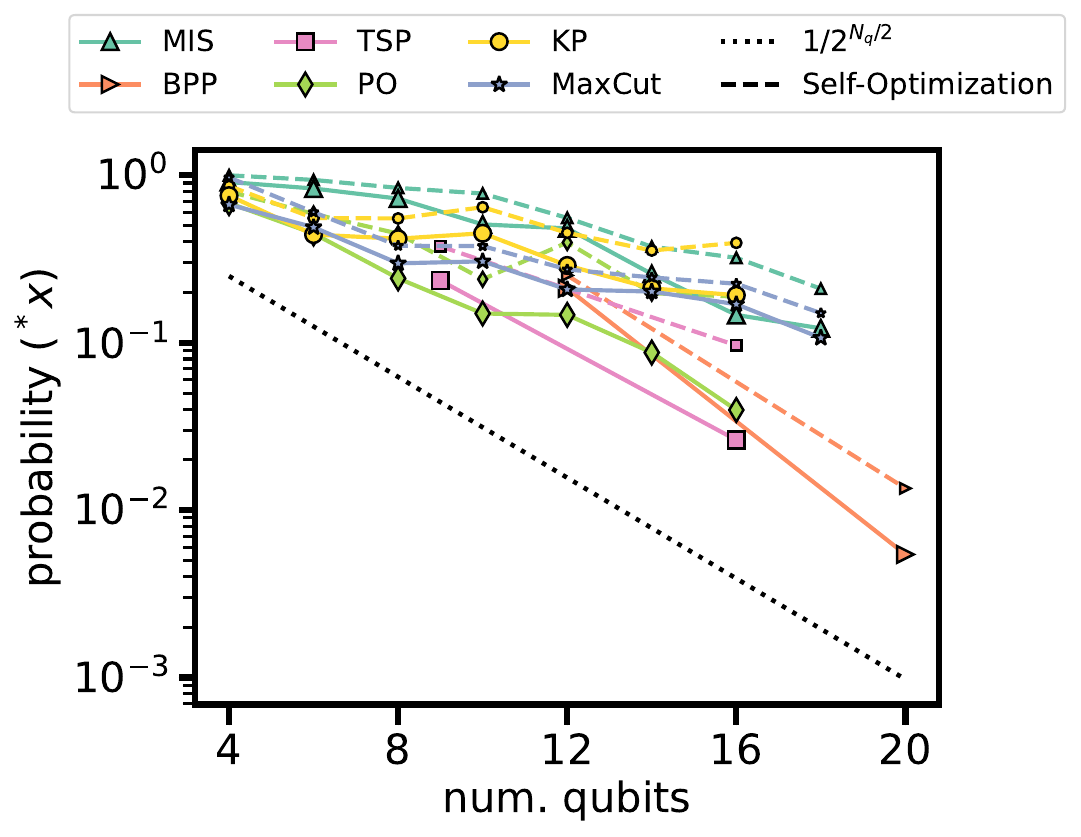}
\caption{\label{tl_from_bpp_opt} Comparison between TL (solid line)  and self-optimization (dashed line) for different COPs (see legend). Here, each marker represents the mean value over 5 random cases. The dashed lines with small markers represent the problems optimized using the procedure in Sec.~\ref{Sec:QUBO}, and the solid lines with big markers represent the results of applying TL from the BPP. The quadratic speedup over random guessing (black dotted line) is presented as a guiding line.}
\end{figure}

Figure~\ref{tl_from_bpp_opt} shows a comparison between TL from a BPP with 3 items vs.~self-optimization for the mean value of the optimal $probability(^*x)$ of the different COPs. Self-optimization refers to the optimization of the $\gamma$ and $\beta$ parameters for each specific problem using COBYLA with a maximum number of iterations given by $max\_iter = 20 p N_q$, where $N_q$ is the number of qubits and $p$ the number of QAOA layers. The election of $max\_iter$ is arbitrary with the main objective of setting a maximum number of iterations equal for problems with the same size. However, it could be beneficial for some of them to have more iterations while for others there is no improvement after a few hundred iterations.

The guiding black dotted line indicates a quadratic speedup over random guessing, i.e., a reduction in the search space to $\mathrm{O}(2^{N_q/2})$. The trend for all problems using QAOA is better than a quadratic speedup over random guessing. Comparing self-optimization and transfer learning, the probability of success is slightly better for the self-optimization method. However, this improvement does not fully compensate for the resources spent in the self-training step. For instance, the number of samples required to observe the optimal solution with a $99\%$ probability, $N_{samples}^{99\%} = \log{(1 - 0.99)}/\log{(1-probability(^*x))}$, for the 14-qubit (18-qubit) MaxCut, is on average $N_{samples}^{99\%} = 16$ (28) for self-optimization and $N_{samples}^{99\%} = 20$ (41) for TL. The largest difference in number of samples is for 20-qubit BPP with $N_{samples}^{99\%} = 339$ for self-optimization and $N_{samples}^{99\%} = 848$ for TL. The number of iterations times the samples per iteration needs to be taken into account to get an estimate of the resources needed.

\begin{figure}[!tbh]
\centering
\includegraphics[width=8.5cm]{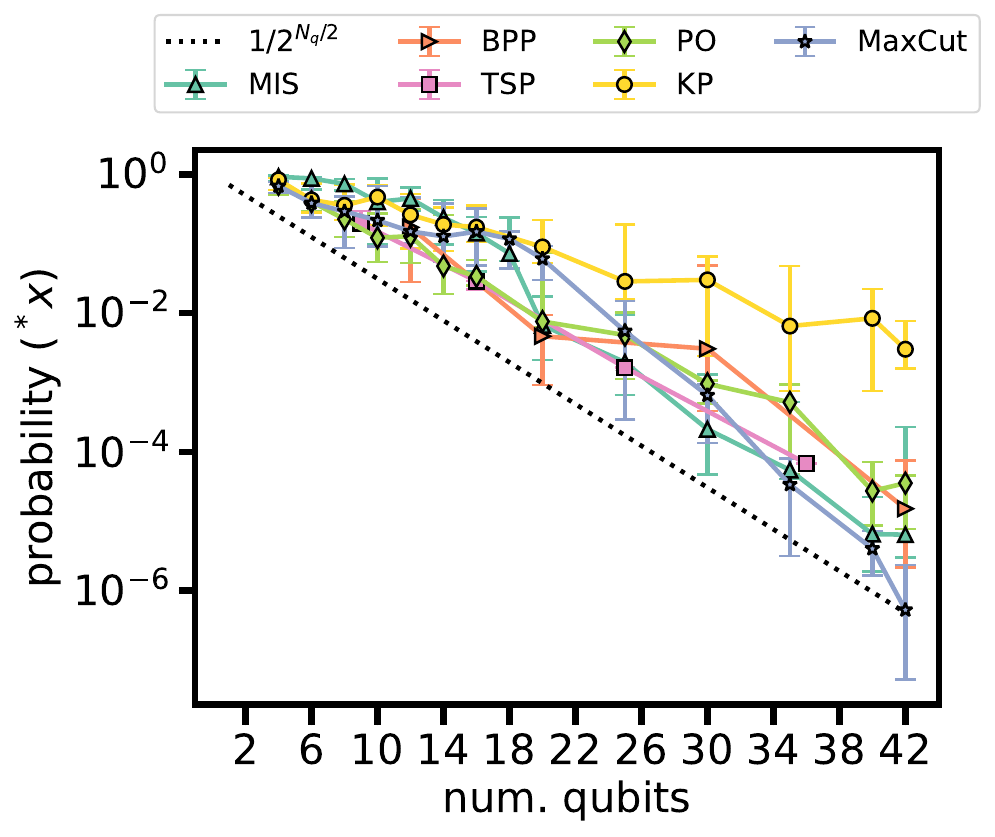}
\caption{\label{tl_bpp} TL from BPP of 3 items to different COPs. Shown is the mean value of 5 random instances of each problem size of the different COPs. Markers represent the mean value. Error bars represent quartiles Q1 and Q3. The guiding line is the same as in Fig.~\ref{tl_from_bpp_opt}.}
\end{figure}

Figure~\ref{tl_bpp} shows the results of applying TL to larger BPP problems with up to 42 qubits, i.e., we take the resulting 12-qubit parameters and apply the same QAOA schedule to significantly larger problems. The mean probability of success is above a quadratic speedup over random guessing for all the problems, which is a good indication of the generalization capabilities of the transferred BPP parameters. The best performance and scaling trend are given by the KP. This scaling is not necessarily related to the TL capabilities because self-optimization shows a similar performance up to the size tested. The improvement could be related to the KP energy structure. The lowest performance is for the MaxCut problem with an error bar going below the quadratic speedup over random guessing. For this problem, the scaling below 18 qubits seems similar to that of the KP, however, as the problem grows in size, the scaling trend changes considerably. It could be related to the maximum values of $\beta$ and $\gamma$, and these values could be rescaled as the problem size grows, similar to the rescaling in \cite{Shaydulin2023b}. The other problems seem to have similar performance that holds as the problem size grows.

\begin{figure}[!tbh]
\centering
\includegraphics[width=8.5cm]{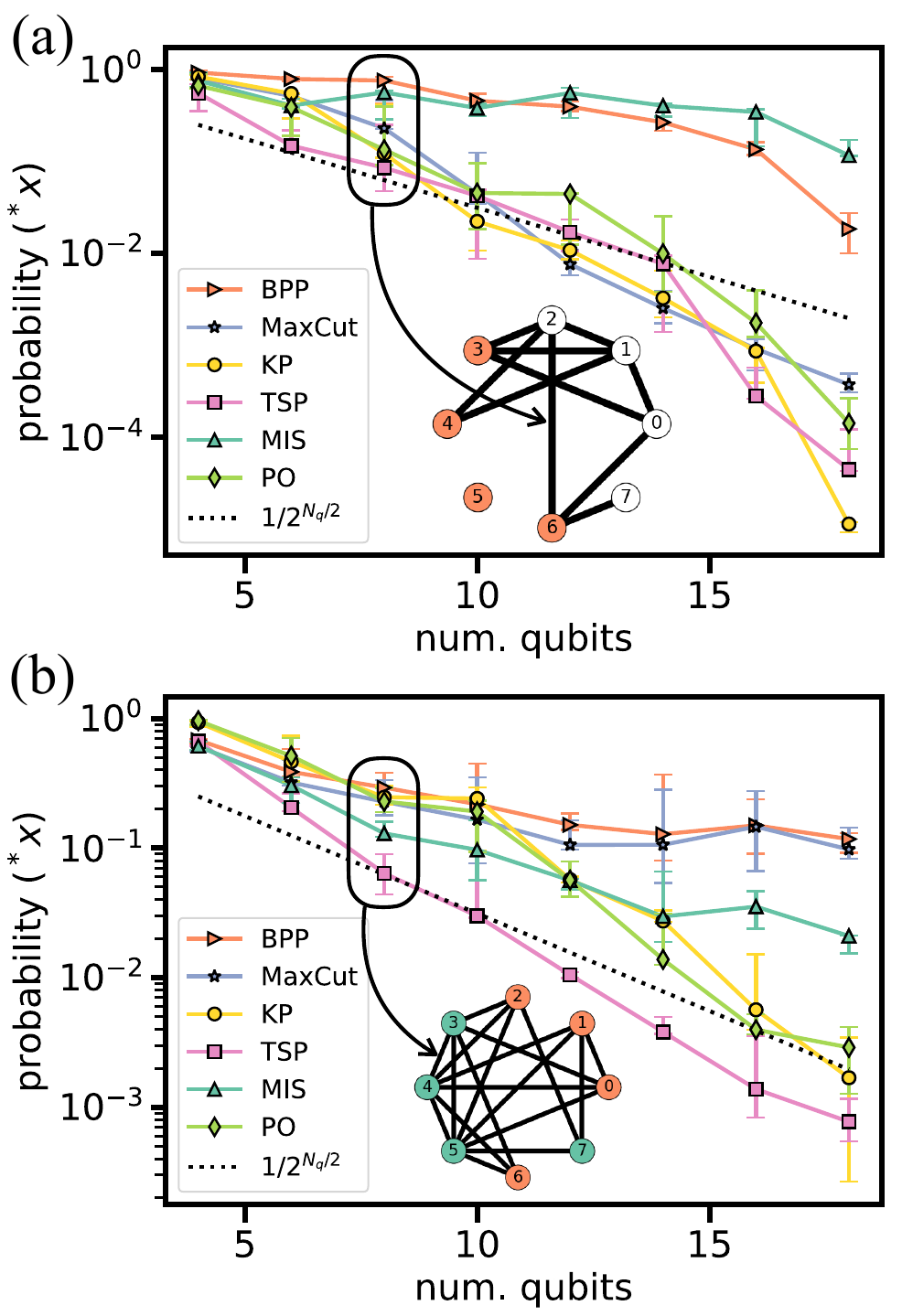}
\caption{\label{tl_mis} TL from different COPs to (a) the MIS problem, (b) the MaxCut problem. Using the $\gamma_p$ and $\beta_p$ parameters from different COPs, 5 random instances for each problem size of the MIS are solved. Markers represent the median value and error bars represent the Q1 and Q3 quartiles. The insets represent one instance of the respective problem solved using the TL parameters of the other COPs.}
\end{figure}

Figure~\ref{tl_mis} shows the results of applying TL from different COPs to the 5 random instances from 4 to 18 qubits of (a) the MIS and (b) MaxCut. In Fig.~\ref{tl_mis} (a), we see that the best performance is obtained by TL parameters from an MIS instance (triangle-up) or a BPP instance (triangle-right). The first can be explained by the fact that different instances of the same problem (MIS) have similar Hamiltonian structures. However, the favorable results for the BPP constitute an interesting empirical observation with no obvious explanation for its generalization capabilities. Figure~\ref{tl_mis}(b) shows that this observation also holds when transferring to the MaxCut problem, i.e., the best performance is obtained by TL from MaxCut to MaxCut (star) or again from BPP to MaxCut (triangle-right).

\subsection*{Solutions on quantum hardware}
Next, we show how TL behaves when executing the QAOA algorithm for $p=10$ on real quantum hardware using the BPP parameters from Fig. \ref{tl_qaoa} (b). The problem used is the MIS for random problems with sizes 8, 14, and 18 qubits. The number of samples used is 1000 for the 8 qubit devices and 5000 for the 14 and 18 qubit devices.

\begin{figure*}[!tbh]
\centering
\includegraphics[width=18cm]{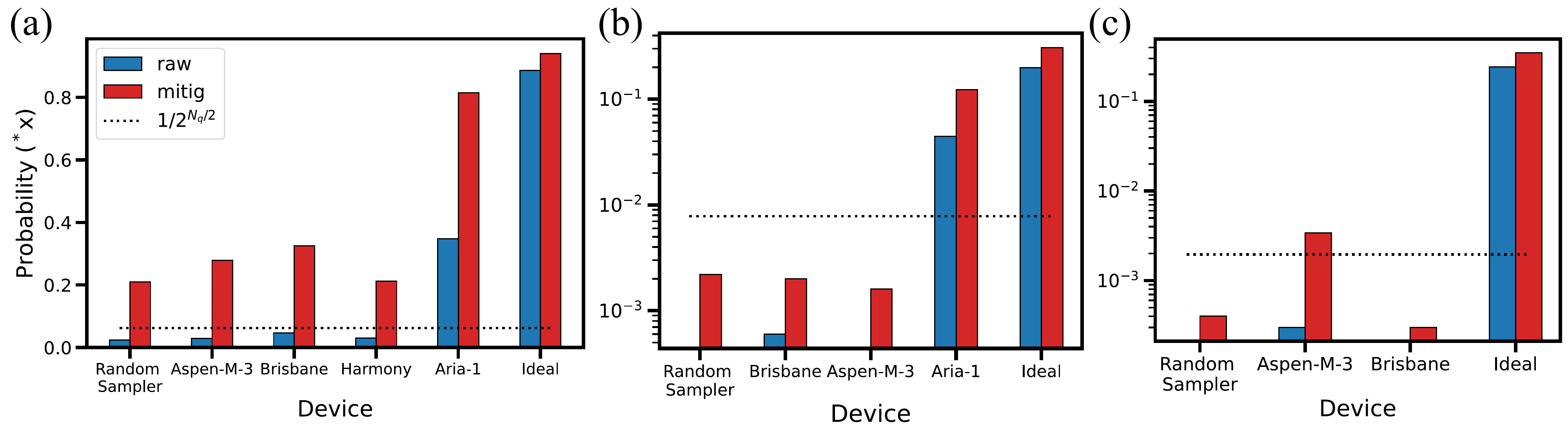}
\caption{\label{real_dev} Success rate to find the optimal solution for different MIS problem sizes using real devices (a) MIS with 8 qubits with 89.1\% of ideal probability (b) MIS with 14 qubits and 19.8\% ideal probability (c) MIS with 18 qubits and  24.1\% ideal probability. The blue bars represent the raw sampling results using the different devices while the red bars are  the percentage of bitstrings with optimal solution and Hamming distance 1 from the optimal solution. The dashed line represents a quadratic improvement in the probability of success compared to random guessing.}
\end{figure*}

In Fig.~\ref{real_dev} (a), we show results for solving an 8-qubit MIS problem on Rigetti's Aspen-M-3 (2.9\%, 27.9\%), IBM's Brisbane (4.7\% raw, 32.5\%), IonQ's Harmony (3\%, 21.2\%), IonQ's Aria-1 (34.8\%, 81.5\%), an ideal simulator (89.1\%, 94.0\%), and random sampling (2.4\%, 21\%). In this problem, there are 240 CNOT gates on a fully connected device. We can see that at these large circuit depths, Aria is the only device with a distribution resembling the ideal case; all others are similar to a random bitstring generator.

In Fig.~\ref{real_dev} (b), we show results for solving a 14-qubit MIS problem on Rigetti's Aspen-M-3 (0.0\%, 0.2\%), IBM's Brisbane (0.06\%, 0.16\%), IonQ's Aria-1 (4.46\%, 12.3\%), the ideal simulator (19.8\%, 30.7\%), and a Random Sampler (0\%, 0.22\%). The probability of connection between different nodes is set to 40\%, and 640 CNOT gates are applied on a fully connected quantum device. Once again, the results indicate that the probability distribution of the ideal case is very different from the Brisbane and Aspen-M-3 results, whereas for Aria the resemblance with the ideal case is still observable.

Finally, in Fig. (\ref{real_dev})-(c) we show results for solving an 18-qubit MIS problem on Rigetti's Aspen-M-3 (0.03\%, 0.34\%), IBM's Brisbane (0.0\%, 0.03\%), the ideal simulator (24.1\%, 34.8\%), and a Random Sampler (0.0\%, 0.04\%). In this case, the probability of connection between different nodes is 40\%, $p=10$, and the problem requires 1020 CNOT gates on a fully connected device (which corresponds to 3657 ECR gates after transpilation to IBM's Brisbane device). Although IonQ's Aria would also have enough qubits to run this case, testing it was prevented by a limitation in the number of 1 and 2-qubit gates to 950 and 650, respectively. In this case, there is a slight improvement in the Aspen-M-3 result compared to the Random Sampler, but it is not significant enough to conclude that partial information about the probability distribution is recoverable.

\subsection*{Cross-platform TL}
\begin{figure}[!tbh]
\centering
\includegraphics[width=8.5cm]{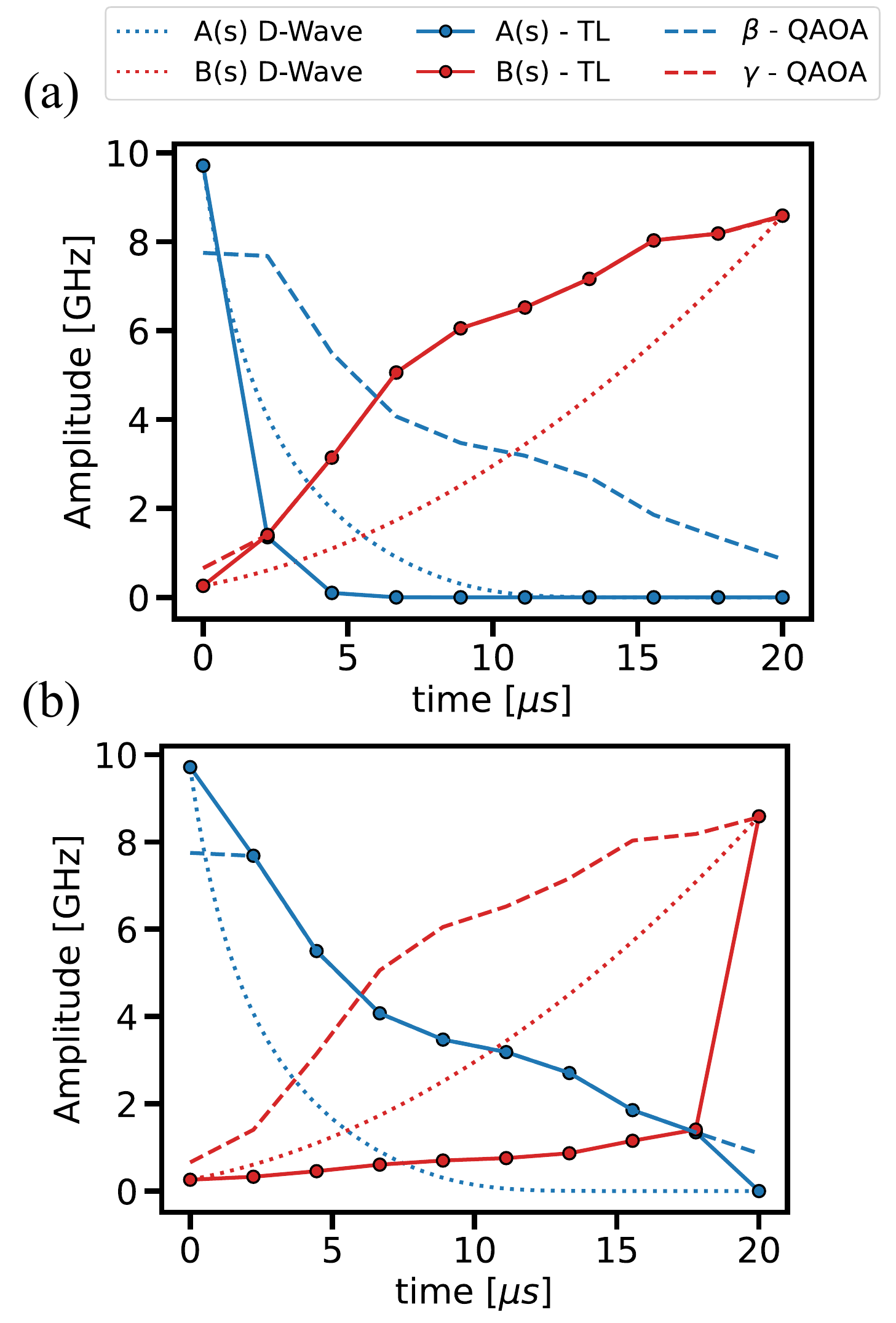}
\caption{\label{tl_to_dwave} Example for cross-platform TL parameters on the D-Wave Advantage protocol. (a) TL to the cost Hamiltonian schedule, and (b) TL to the mixer Hamiltonian schedule.  The dotted line represents the default schedule used by D-Wave, the solid line is the new schedule with TL applied, and the dashed line reflects the BPP-QAOA schedule. The blue (red) color represents the mixer (cost) Hamiltonian protocol.}
\end{figure}

In this section, we study cross-platform TL by transferring the $\gamma$ and $\beta$ parameters of the 3 items BPP that have shown good TL capabilities to a different quantum computing platform, namely a D-Wave Advantage quantum annealer. Figure~\ref{tl_to_dwave} shows the two modified annealing schedules for $A(s)$ and $B(s)$, respectively. Note that we modify D-Wave Advantage $A(s)$ and $B(s)$ schedules separately because the device does not allow the modification of both of them at the same time. Fig.~\ref{tl_to_dwave}-(a) shows the TL of the $\gamma$ parameters of the BPP to the schedule $B(s)$ in D-Wave Advantage. Fig.~\ref{tl_to_dwave}-(b) shows the TL of the $\beta$ parameters of the BPP to the schedule $A(s)$ in D-Wave Advantage. A further restriction in the D-Wave Advantage schedule is that the initial and final annealing points must correspond to the $A(s)$ and $B(s)$ points.

\begin{figure}[!tbh]
\centering
\includegraphics[width=8cm]{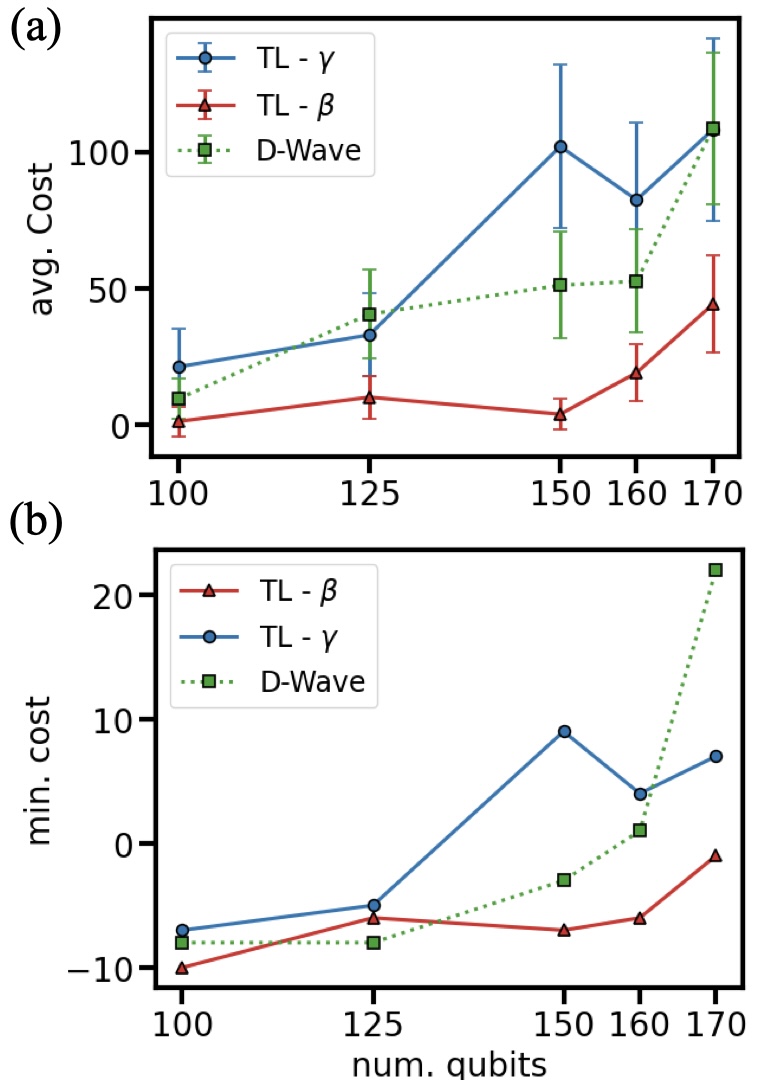}
\caption{\label{tl_dwave_results} Cross-platform TL for the MIS using BPP parameters (a) average value using 5000 samples on D-Wave Advantage (b) minimum cost of the 5000 samples.}
\end{figure}

 Figure.~\ref{tl_dwave_results} shows the TL to the MIS problem for problem sizes 100, 125, 150, 160, and 170 qubits. The results are obtained using 5000 samples for each problem size. Equation \ref{QUBO_mis} is the QUBO formulation used to implement the problem and to evaluate the cost function. The green dotted lines are the results using the D-Wave default annealing schedule, the red triangle line is the TL of the mixer Hamiltonian parameters, and the blue circle line is the TL of the cost Hamiltonian parameters. Figure (\ref{tl_dwave_results})-(a) shows the mean cost with error bars representing the standard deviation of the 5000 sample costs. Figure (\ref{tl_dwave_results})-(b) shows the minimum cost of the 5000 samples for each problem size. These results show a consistent improvement in the distribution of solutions using TL of $\beta$ QAOA parameters to the $A(s)$ parameters of the mixer Hamiltonian schedule of D-Wave Advantage both in terms of average and minimum value.

\section{Conclusions}\label{Sec:Conclusion}
We have presented transfer learning (TL) of QAOA parameters in the context of COP, a methodology that involves using pre-optimized QAOA parameters to solve different COPs. This method therefore does not require extra steps of classical optimization, but further parameters optimization using the TL parameters as a starting point could be beneficial. We show that the BPP has great generalization capabilities, i.e., the parameters for small instances of BPP are good for larger instances of the same problem and effective for instances of other COPs. The successful TL between instances of BPP to different COPs can be intuitively pictured as the consequence of optimizing an effective, digitized annealing schedule that works within the limit of a small number of layers (p = 10). This is the opposite of what happens with other COPs, where the parameters found do not perform well on other problems.

We test TL using KP, BPP, MIS, MaxCut, TSP, and PO. First, we do self-optimization for these problems, exploring random instances with up to 20 qubits. Then, we select the case with the best performance to find optimal solutions to the other problems. In our case, the parameters used are those of the BPP for 3 items (12 qubits). We use those parameters in different instances of the same and other problems for up to 42 qubits and find that for all of them, the probability of finding the ground state is above the quadratic speedup over random guessing. This suggests that there are $\beta$ and $\gamma$ parameters that generalize well over different COPs in their QUBO formulation.

Next, we show that coherent outputs for problem sizes up to 14 qubits are still present in current quantum technology. We test 3 different instances of the MIS problem with 8, 14, and 18 qubits with the devices used being Rigetti's Aspen-M-3, IBM's Brisbane, IonQ's Harmony, and Aria. In the case of 8 qubits, a direct comparison between the two generations of IonQ technology is possible, Harmony with a success probability of 3.0$\% $ and Aria 34.8$\%$. This means an order of magnitude improvement between these two generations of IonQ trapped ions for sampling optimal solutions. There is still room for improvement and benchmarking with this methodology is a promising tool to see the evolution of quantum technology for sampling.

Though self-optimization gives a better probability$(^*x)$ than TL for every COP, this improvement does not fully compensate for the resources needed for the optimization loop needed to find the parameters. TL is also convenient for the classical simulation of large problems that typical computers cannot handle ($N_q > 30$). Compared to the self-optimization method it only requires one iteration compared to the hundreds of self-optimization. In the case of $N_q > 35$, only high-performance computing (HPC) systems can handle the simulations, making self-optimization computationally too expensive to even be considered. Self-optimization issues are more notorious in real quantum hardware where the cost of finding QAOA parameters is in terms of the number of samples and circuits used and classical solvers find it hard to work with noisy quantum hardware \cite{Bonet-Monroig2023}. This makes TL a useful alternative that can compensate for the problems associated with the classical optimization loop even if TL parameters are not optimal.

Finally, we show that cross-platform TL is possible. We use D-Wave Advantage to test MIS problems between 100 and 170 qubits using the QAOA parameters of the BPP, Fig.~\ref{tl_qaoa}. Two cases are tested, one with the modification of the mixer Hamiltonian B(s) and one with the modification of the cost Hamiltonian annealing protocol A(s). We find a consistent improvement in terms of the minimum and average cost using the mixer Hamiltonian modified schedule.

 A future direction could be to extended the analysis of transfer learning in the context of Higher-Order Unconstrained Binary Optimization (HUBO) formulations of COPs, e.g., the one presented in \cite{Glos2022}.

\begin{acknowledgments}
\vspace{-10pt}
J. A. Montanez-Barrera acknowledges Unitary Fund for their support in the development of the combinatorial optimization library used in this work within OpenQAOA SDK. We acknowledge AWS Amazon Braket for the time used across the quantum devices Aspen-M3, Harmony, and Aria-1. We acknowledge the use of IBM Quantum services for this work. The views expressed are those of the authors, and do not reflect the official policy or position of IBM or the IBM Quantum team. J. A. Montanez-Barrera acknowledges support by the German Federal Ministry of Education and Research (BMBF), funding program Quantum Technologies - from basic research to market, project QSolid (Grant No. 13N16149).
D.~Willsch acknowledges support from the project J\"ulich UNified Infrastructure for Quantum computing (JUNIQ) that has received funding from the German Federal Ministry of Education and Research (BMBF) and the Ministry of Culture and Science of the State of North Rhine-Westphalia.
The authors gratefully acknowledge the Gauss Centre for Supercomputing e.V. (www.gauss-centre.eu) for funding this project by providing computing time on the GCS Supercomputer JUWELS at Jülich Supercomputing Centre (JSC).

\end{acknowledgments}
\clearpage  

\bibliography{References}

\clearpage  
\onecolumngrid  
\appendix

\section{Supplementary Material}\label{appendix}
In this section, a description of the different problems used in this work is presented. For each problem, the constraints are encoded using penalization terms. We use squared penalty terms for equality constraints, and the unbalanced penalization approach \cite{Montanez-Barrera2022} for inequality constraints.

\subsection{QUBO Formulation}
One approach to represent combinatorial optimization problems is through the Quadratic Unconstrained Binary Optimization (QUBO) formulation. The QUBO formulation expresses the problem as a quadratic objective function that depends on binary variables. The objective is to minimize this function by determining the values of the binary variables, subject to certain constraints. Combinatorial problems that can be represented by the QUBO formulation have functions of the form

\begin{equation}
f(\mathbf{x}) = \sum_{i=0}^{n-1} \sum_{j=0}^{n-1} q_{ij} x_i x_j.
\end{equation}
Here, $n$ represents the number of variables, $q_{ij} \in \mathbb{R}$ are coefficients associated with the specific problem and $x_i \in \{0,1\}$ are the binary variables. It is important to note that in this formulation, $x_i x_i \equiv x_i$ and $q_{ij} = q_{ji}$.

A general form of a combinatorial optimization problem that can be solved using Quantum Processing Units (QPUs) is characterized by a cost function

\begin{equation}\label{QUBO_form}
f(\mathbf{x}) = 2\sum_{i=0}^{n-1} \sum_{j > i} q_{ij}x_i x_j + \sum_{i=0}^{n-1} q_{ii} x_i, 
\end{equation}
and additionally, linear equality constraints are given by

\begin{equation}
\sum_i c_i x_i = C, \quad c_i \in \mathbb{Z},
\end{equation}
and linear inequality constraints are given by

\begin{equation}\label{Eq:ineq}
\sum_i w_i x_i \leq W, \quad w_i \in \mathbb{Z},
\end{equation}
can be added. Here, $q_{ij}$, $c_i$, and $w_i$ are parameters of the problem. To transform problems with constraints into the QUBO formulation, the constraints are usually incorporated as penalization terms. The equality constraints are included in the cost function using a penalization term of the form

\begin{equation}
\lambda_0 \left(\sum_i c_i x_i - C\right)^2.
\end{equation}
Here, $\lambda_0$ is a penalization coefficient that should be chosen appropriately to obtain sufficient solutions that satisfy the equality constraint, and $C$ is a constant value given by the constraint.

For the inequality constraints, we use the \textit{unbalanced penalization} \cite{Montanez-Barrera2022} encoding which is a heuristic method for including inequality constraints as penalization terms in the QUBO formulation of combinatorial optimization problems. The method has been shown to outperform the slack variables encoding for the TSP, BPP, KP, and collateral optimization \cite{Montanez-Barrera2022, Montanez-Barrera2023, Giron2023}. Starting from Eq.(\ref{Eq:ineq}), the method adds a penalization term $\xi(\mathrm{x})$ to the objective function given by

\begin{equation}
\xi(\mathrm{x}) = -\lambda_1 h(\mathrm{x}) + \lambda_2 h(\mathrm{x})^2,
\label{eq:xi}
\end{equation}
where $h(\mathrm{x}) = W - \sum_i w_i x_i$ and $\lambda_{1,2}$ are penalization coefficients that should be chosen to guarantee that the constraint is fulfilled. The term $\xi(\mathrm{x})$ is unbalanced, meaning that it imposes a larger penalization for negative values of $h(\mathrm{x})$ (i.e., when the constraint is not satisfied) than for positive values. The QUBO formulation using the unbalanced penalization approach is given by 

 \begin{equation}\label{QUBO_unbalanced}
 \min_x \left(2\sum_{i, j>i} q_{ij}x_{i}x_{j} + \sum_{i}q_{ii} x_i + \lambda_0 \left(\sum_{i} c_i x_i - C\right)^2 - \lambda_1h(x) +  \lambda_2 h(x)^2 \right).
 \end{equation}

The parameters for the different problems studied in this work are shown in Table \ref{Table1}. In general, this method does not guarantee that the optimal solution is encoded in the ground state of the Ising Hamiltonian. For the probability of the BPP, PO, and KP in Figs.~\ref{tl_from_bpp_opt} and \ref{tl_bpp},  we choose as optimal solution the ground state of the new Ising Hamiltonian that in the majority of the cases is the optimal solution of the original problem. The last step to represent the QUBO problem as an Ising Hamiltonian is to change the $x_i$ variables to spin variables $z_i \in \{1, -1\}$ by the transformation $x_i = (1 - z_i) / 2$. Note that Eq.(\ref{QUBO_unbalanced}) can ultimately be rewritten as Eq.(\ref{QUBO_form}) plus a constant value. Hence, Eq.(\ref{QUBO_form}) represented in terms of the Ising model reads

\begin{equation}\label{IsingH}
H_c(\mathrm{z}) = \sum_{i=0}^{n-1}\sum_{j>i}^{n-1} J_{ij} z_i z_j + \sum_{i=0}^{n-1} h_{i} z_i + \text{offset},
\end{equation} 
where $J_{ij}$ and $h_i$ are real coefficients that represent the combinatorial optimization problem, and the offset is a constant value. Since the offset does not affect the location of the optimal solution, it can be left out for the sake of simplicity. In the following subsection, the COPs are presented. These problems can be translated into the Ising Hamiltonian representation following the methodology presented in this section. The last step we use to solve the problems using QAOA is to normalize the Hamiltonian by the maximum weight in the Hamiltonian, i.e., $\max\{J_{ij}, h_i\}$.

\begin{table}[ht]
\caption{\label{Table1}Parameters $\lambda_{0,1,2}$ for the TSP, BPP, KP, PO, and MIS used to translate the combinatorial optimization problems into the QUBO representation using the unbalanced penalization approach (Eq.~(\ref{QUBO_unbalanced})). For all  equality constraints of each problem, we use the same $\lambda_0$, and for the inequality constraints the same $\lambda_{1,2}$.}
\begin{center}
\begin{tabular}{c c c c}\hline
& $\lambda_0$& $\lambda_1$ & $\lambda_2$ \\ 
 \hline
TSP & 23 & - & -\\
BPP & 15 & 4.2 & 0.4\\
KP & - & 0.96 & 0.04\\
PO & - & 0.97 & 0.06\\
MIS & - & 1 & 1 \\
 \hline
\end{tabular}
\end{center}
\end{table}

\subsection{Combinatorial optimization problems}
\subsubsection{Traveling salesman problem}
The TSP is a well-known combinatorial optimization problem that aims to determine the shortest possible route to visit a given set of cities and return to the starting city. This problem has various practical applications, including route planning, circuit board drilling, and DNA sequencing. 
A QUBO formulation of the TSP can be obtained using a time encoding of the route that the traveler passes on a Hamiltonian cycle \cite{Lucas2014}. For the asymmetric and symmetric forms, this TSP formulation requires $n^2$ variables for $n$ cities (we note that in principle, one can reduce this to $(n-1)^2$ variables by fixing the starting point). It needs $2n$ equality constraints and avoids complications associated with sub-tours. The TSP formulation is given by

\begin{equation}\label{TSP-QP}
\min\sum_{t=0}^{n-1}\sum_{i=0}^{n-1}\sum_{j \neq i, j=0}^{n-1} c_{ij}x_{i,t}x_{j,t+1},
\end{equation}
subject to the set of constraints,

\begin{equation}\label{TSPEQ1}
\sum_{i=0}^{n-1} x_{i, t} = 1 \qquad \forall t = 0, ..., n-1,
\end{equation}
and 
\begin{equation}\label{TSPEQ2}
\sum_{t=0}^{n-1} x_{i, t} = 1 \qquad \forall i = 0, ..., n-1.
\end{equation}
Equation (\ref{TSPEQ1}) expresses that at every time $t$ exactly one city is visited, while Eq.(\ref{TSPEQ2}) expresses that every city $i$ is visited at exactly one time. For this problem, we use instances with 3, 4, 5, and 6 cities, where the distances between cities $c_{ij}$ are randomly chosen from a normal distribution with a mean equal to 10 and a standard deviation equal to 0.1.

\subsubsection{Knapsack Problem}
The KP involves selecting a subset of items from a larger set, each with a certain weight $w_i$ and value $v_i$, in such a way that the total weight does not exceed a given limit while maximizing the total value. Formally, the cost function is 

\begin{equation}\label{knapsack_cost}
    \max \sum_{i=0}^{n-1} v_i x_i, 
\end{equation}
subject to the single inequality constraint,

\begin{equation}\label{knapsack_inequality}
    \sum_{i=0}^{n-1} w_i x_i \leq W,
\end{equation}
where $x_i = 1, 0$ indicates that an item is included or not, and $W$ is the maximum weight. We select item values $v_i$ ranging from 5 to 63 randomly, weights $w_i$ from 1 to 20 randomly, and maximum weight $W = \frac{1}{2}\sum_i w_i$.

\subsubsection{Portfolio Optimization}
The goal of PO is to create a balanced portfolio out of a selection of financial assets, which should maximize future returns while taking into account the total risk of the investment. The information we have about the assets is their past returns $\mu_i$ and the covariances between assets $\sigma_{ij}$, with which the problem can be formulated as follows

\begin{equation}\label{portfolio_cost}
    \max \sum_{i=0}^{n-1} \mu_i x_i - q \sum_{i=0}^{n-1} \sigma_{ij} x_i x_j,
\end{equation}
subject to the inequality constraint

\begin{equation}\label{portfolio_inequality}
    \sum_{i=0}^{n-1} c_i x_i \leq B,
\end{equation}
where (similar to the KP), the $x_i$ indicates whether an asset is selected as part of the portfolio or not, and $B$ is the total budget. The factor $q$ controls how much risk is taken. If it is small, the second term in Eq.(\ref{portfolio_cost}) becomes negligible and the returns $\mu_i$ will be dominant in determining the optimal solutions. In the main text, problem sizes ranging from 4 to 42 are presented. The values of the expected return $\mu_i$ are randomly chosen between 0 and 1. The correlation matrix $\sigma_{i,j}$ is selected randomly from the set $[-0.1, 0, 0.1, 0.2]$. Asset costs $c_i$ are randomly chosen between 0.5 and 1.5. The budget is set as $B = \frac{1}{2} \sum_i c_i$.

\subsubsection{Maximal Independent Set}
The MIS problem asks to find the largest subset of vertices of a graph, such that no two vertices in the subset are adjacent. This subset is then called independent. Formally, for an undirected graph $G = (V,E)$, the problem formulation is 

\begin{equation}\label{MIS_cost}
    \max \sum_{v \in V} x_v,
\end{equation}
subject to

\begin{equation}\label{MIS_inequality}
    x_u + x_v \leq 1 \qquad \forall (u, v) \in E,
\end{equation}
where the binary variable $x_v$ determines whether a vertex is included in the subset or not. We select problem sizes between 4 and 42 qubits and the probability of having an edge between nodes of 50\%. The constraints in this case are added to the QUBO formulation as $2x_ix_j$ if an edge is present. Therefore, the QUBO formulation for this problem is given by 

\begin{equation}\label{QUBO_mis}
    \min \sum_{i \in V} - x_i + \sum_{i,j \in E} 2 x_i x_j.
\end{equation}

\subsubsection{Bin Packing Problem}
The BPP involves the efficient packing of a collection of items into the minimum number of bins, where each item has an associated weight and the bins have a maximum weight capacity. This problem finds applications in various real-world scenarios like container scheduling \cite{Yan2022Book}, and FPGA chip design \cite{Kroes2020} among others. The BPP is classified as an NP-hard problem due to its computational complexity. The problem can be formulated as follows, minimize the total number of bins used given by the objective function

\begin{equation}\label{BPPmin}
\min \sum_{j=0}^{m-1} y_j,
\end{equation}
subject to the following constraints. Each bin's weight capacity should not be exceeded

\begin{equation}\label{BPPIC}
\sum_{i=0}^{n-1} w_i x_{ij} \le W y_j \quad \forall j=0,...,m-1,
\end{equation}
and each item can only be assigned to one bin

\begin{equation} \label{BPPEC}
\sum_{j=0}^{m-1} x_{ij} = 1 \quad \forall i = 0, ..., n-1.
\end{equation}
Binary variables indicating item-bin assignments and bin utilization

\begin{equation}\label{BPPB1}
x_{ij} \in {0,1} \quad \forall i=0,..,n-1 \quad \forall j=0,..,m-1,
\end{equation}
\begin{equation}\label{BPPB2}
y_j \in {0,1} \quad \forall j=0,..,m-1.
\end{equation}

In the above equations, $n$ represents the number of items (nodes), $m$ represents the number of bins, $w_{i}$ is the weight of the $i$-th item, $W$ denotes the maximum weight capacity of each bin, and $x_{ij}$ and $y_j$ are binary variables representing the presence of item $i$ in bin $j$ and the use of bin $j$, respectively. The objective function in Eq.(\ref{BPPmin}) aims to minimize the number of bins used, while Eq.(\ref{BPPIC}) enforces the constraint on bin weight capacity. Eq.(\ref{BPPEC}) ensures that each item is assigned to only one bin, and Eqs. (\ref{BPPB1}) and (\ref{BPPB2}) define the binary nature of variables $x_{ij}$ and $y_j$. The main text considers scenarios involving 3, 4, 5, and 6 items. The weight of each item $w_i$ is randomly chosen from 1 to 10, and 20 is the maximum weight $W$ of the bins. The Lagrange multipliers $\lambda_{0,1,2}$ in Eq.(\ref{QUBO_unbalanced}) for this problem are 15, 4.2, and 0.4, respectively.

\subsubsection{Maximum Cut}
The MaxCut problem involves determining the partition of the vertices in an undirected graph such that the total weight of the edges between the two sets is maximized. For an undirected graph \( G = (V, E) \), the problem is formulated as

\begin{equation}\label{MaxCut_cost}
    \max \sum_{(i, j) \in E} w_{ij} (x_i + x_j - 2 x_i x_j),
\end{equation}

where \( w_{ij} \) represents the weight of the edge between vertices \( i \) and \( j \), and \( x_i \) and \( x_j \) are binary variables that determine the partition of vertices. The goal is to maximize the sum of edge weights over all edges in the cut. The binary variables \( x_i \) and \( x_j \) take values of 0 or 1, indicating the membership of vertices in different sets of the partition. If \( x_i \) and \( x_j \) are different, the edge contributes to the objective function. 
\end{document}